\documentclass[prb,twocolumn,superscriptaddress]{revtex4-1}

\usepackage{amsmath,color,amsfonts,graphicx,bm}

\newcommand{\ket}[1]{\vert#1\rangle}

\begin{document}
\title{Semionic resonating valence bond states}
\author{Mohsin Iqbal} 
\affiliation{JARA Institute for Quantum Information, RWTH Aachen, 52056 Aachen, Germany}
\author{Didier Poilblanc}
\affiliation{Laboratoire de Physique Th\'eorique, C.N.R.S.\ 
    and Universit\'e de Toulouse, 31062 Toulouse, France}
\author{Norbert Schuch}
\affiliation{JARA Institute for Quantum Information, RWTH Aachen, 52056 Aachen, Germany}

\begin{abstract}
The nature of the kagome Heisenberg antiferromagnet (HAFM) is under
ongoing debate. While recent evidence points towards a $\mathbb
Z_2$ topological spin liquid, the exact nature of the topological phase is
still unclear.  In this paper, we introduce semionic Resonating Valence
Bond (RVB) states, this is, Resonating Valence Bond states which are in
the $\mathbb Z_2$ ordered double-semion phase, and study them using Projected
Entangled Pair States (PEPS). We investigate their physics and study 
their suitability  as an ansatz for the HAFM, as compared to 
a conventional RVB state which is in the Toric Code $\mathbb Z_2$
topological phase. In particular, we find that a suitably optimized
``semionic simplex RVB`` outperforms the equally optimized conventional
''simplex RVB`` state, and that the entanglement spectrum (ES) of the
semionic RVB behaves very differently from the ES of the conventional RVB,
which suggests to use the ES to discriminate the two phases. Finally, we
also discuss the possible relevance of space group symmetry breaking in
valence bond wavefunctions with double-semion topological order.  
\end{abstract} 

\maketitle

\section{Introduction}

Strongly correlated quantum systems exhibit a wide range of exciting and
unconventional physical phenomena.  One such effect which has recently
received much attention due to its exotic properties are topological spin
liquids, which---despite strong antiferromagnetic interactions---do not
order magnetically even at very low temperatures due to the presence of
strong quantum fluctuations; at the same time, these systems order
topologically, which gives e.g.\ rise to fractionalized excitations with anyonic
statistics.~\cite{balents:spin-liquid-review-2010} It is believed that
such phases are realized in materials such as Herbertsmithite, which show
no sign of magnetic ordering down to very low
temperatures,\cite{mendels:herbertsmithite-no-ordering,helton:herbertsmithite-no-symbreaking}
and where indeed recently signatures of fractionalized excitations have been
experimentally observed.~\cite{han:herbertsmithite-anyons}

From a theoretical point of view, a prime candidate model for a
topological quantum spin liquid is the Heisenberg antiferromagnet (HAFM)
on the kagome lattice, which is believed to provide a good approximation
of the physics of Herbertsmithite. Understanding the low-temperature phase
diagram of the kagome HAFM has proven notoriously difficult, with several
different phases competing with each other; the currently most convincing
data, obtained using
DMRG,~\cite{yan:heisenberg-kagome,depenbrock:kagome-heisenberg-dmrg}
suggests that the ground state of the kagome HAFM is a gapped $\mathbb
Z_2$ topological spin liquid, i.e., a spin liquid with a $\log(2)$
correction to the topological entropy. 

Given the hardness of understanding the exact ground state of the kagome
HAFM, in order to obtain a qualitative understanding of the physics of the
kagome HAFM, variational wavefunctions have been considered which are
constructed to capture the physics of the antiferromagnetic interactions.
Resonating Valence Bond (RVB) states have been proposed as
an ansatz for antiferromagnets,~\cite{anderson:rvb} and have helped to
understand the spin liquid nature of antiferromagnets e.g.\ on the square
or kagome lattice; in particular, it
has been found that the kagome RVB state is a $\mathbb Z_2$ topological
spin liquid in the phase of the Toric Code model.~\cite{schuch:rvb-kagome}
Combining this with the observation of $\mathbb Z_2$ topological order in
the aforementioned numerical simulations, this suggests that the kagome
HAFM is a spin liquid in the same phase as the Toric Code.  However, there
is a another topological phase with a topological entropy $\log(2)$,
corresponding to the so-called double semion model; and while these two
phases can in principle be distinguished by their quasi-particle
excitations, this has up to now not been achieved in DMRG calculations. 

In this paper, we apply variational wavefunctions to better understand the
topological nature of the kagome HAFM. To this end, we introduce
\emph{semionic RVB states}, this is, RVB states which are in the phase of
the double semion model, and investigate their properties and their
suitability as a variational ansatz for the kagome HAFM.  We show that
semionic RVB states are in the same phase as the double semion model, and
that they do not exhibit magnetic ordering.  While the energy of the
semionic RVB as an ansatz for the kagome HAFM is not competitive with the
conventional RVB, we find that a two-parameter generalization termed
\emph{semionic simplex RVB} has a variational energy which is even below the
energy of the analogous family of conventional simplex RVB states. We then
proceed to study the entanglement properties of the semionic RVB state,
and find that its entanglement spectrum exhibits features which are
clearly distinct from the entanglement spectrum of the conventional RVB
state (in particular, the minimum in the dispersion is at different
momentum), and which therefore might be useful to discriminate
the two $\mathbb Z_2$ topological phases e.g.\ in DMRG simulations.
An interesting property of our wavefunction is that it explicitly breaks
translational symmetry.  While this might be an artefact of the
construction, we find evidence that symmetry breaking is in fact
energetically favorable, and we discuss possible consequences for the
ground state of the kagome HAFM.

A central tool in our investigation are Projected Entangled Pair States
(PEPS). PEPS allow for an exact description of a variety of states, in
particular renormalization fixed points of topological
phases,~\cite{verstraete:comp-power-of-peps,buerschaper:stringnet-peps,gu:stringnet-peps}
but also Resonating Valence Bond
states,~\cite{verstraete:comp-power-of-peps} where they have allowed for
both analytical and high-precision numerical study of their
properties.~\cite{schuch:rvb-kagome,wang:rvb-square-lattice} 
By combining elements of the PEPS constructions for the double semion model
and the RVB state, we obtain a PEPS description of semionic RVB states,
which we subsequently apply to study their physics.

This paper is structured as follows.  In Sec.~\ref{sec:definitions}, we
introduce semionic RVB states through a mapping to loop models. In
Sec.~\ref{sec:peps-formalism}, we show how to represent semionic RVB
states as PEPS. Finally, in Sec.~\ref{sec:results}, we apply the PEPS
representation of semionic RVBs to study their physics, including their
generalization to semionic simplex RVBs in
Sec.~\ref{sec:subsec:simplex-rvb}.

\section{Semionic RVB states
\label{sec:definitions}
}

\begin{figure}[t]
 \includegraphics[width=\columnwidth]{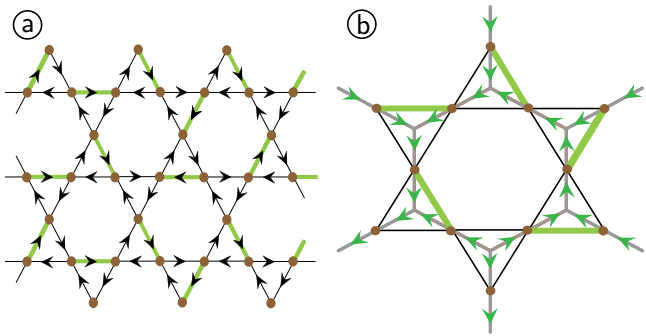}
\caption{
\label{fig:kagomes}
\textbf{(a)} Dimer covering of the kagome lattice. The dimers are
indicated in green, the arrows denote the canonical orientation of the
singlets. \textbf{(b)} Arrow representation of dimer configuration. Each
edge of the dual hexagonal lattice is assigned an orientation (arrow)
which points into the triangle in which the dimer associated with the
corresponding vertex of the kagome lattice lies.
} \end{figure}

In this section, we introduce semionic RVB states.  To this end, consider
the kagome lattice, Fig.~\ref{fig:kagomes}. A \emph{dimer covering} $D$ of
the kagome lattice is a covering of edges with \emph{dimers}, such that
every vertex is adjacent to exactly one dimer, as shown in
Fig.\ref{fig:kagomes}a. We now place spin-$\tfrac12$ particles at the
vertices, and associate to every dimer covering $D$ a state
$\ket{\sigma(D)} = \bigotimes (\ket{01}-\ket{10})$, where the singlets are
placed on the dimers and oriented according to some convention (such as
clock-wise around triangles). The (short-range) RVB state is then given as
\[
\ket{\psi_\mathrm{RVB}} = \sum_D \ket{\sigma(D)}\ .
\]
Analogously, one can define a \emph{dimer model} by replacing
$\ket{\sigma(D)}$ by an orthonormal basis $\ket{D}$ of the space of dimer
coverings $D$. Dimer models have been studied extensively, and it has in
particular been shown that they exhibit $\mathbb Z_2$ topological order
and appear as ground states of local Hamiltonians which are locally
equivalent to the Toric Code
model;~\cite{rokhsar:dimer-models,moessner:dimer-triangular,misguich:dimer-kagome}
more recently, the same could be shown for RVB
states.~\cite{seidel:kagome,schuch:rvb-kagome,zhou:rvb-parent-onestar}

\begin{figure}
\includegraphics[width=\columnwidth]{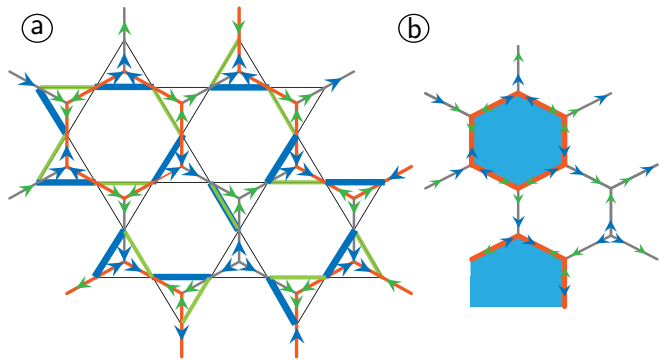}
\caption{
\label{fig:looppatterns}
\textbf{(a)} Mapping between dimer configurations and loop patterns. Given
a reference dimer configuration (blue), any other dimer configuration
(green) can be described by the arrows which need to be flipped in the
corresponding arrow configuration (cf.~Fig.~\ref{fig:kagomes}b). Due to
parity constraints, the edges with a flipped arrow form closed loops on the
honeycomb lattice (red). \textbf{(b)} Loop patterns can be equivalently
described by assigning binary variables (``colors'') to the plaquettes,
and putting strings between different colors.
}
\end{figure}

Dimer coverings of the kagome lattice are in one-to-one correspondence to
loop patterns on the hexagonal lattice.  This can be shown using the
``arrow representation'' introduced by Elser and
Zeng.~\cite{elser:rvb-arrow-representation} To
this end, consider the honeycomb lattice dual to the kagome lattice,
cf.~Fig.~\ref{fig:kagomes}b.  On every vertex of the kagome lattice (this
is, every edge of the hexagonal lattice), we now place an arrow which
points into the triangle in which the dimer adjacent to this vertex lies,
as shown in Fig.~\ref{fig:kagomes}b. By construction, the number of arrows
pointing into any given triangle must be odd. We now fix a ``reference
dimer configuration'' $R$, corresponding to a ``reference arrow
orientation''.  Any other dimer configuration $D$ is now characterized by
the arrows which need to be flipped as compared to the reference arrow
orientation. Due to parity constraints, the number of flipped arrows in
each triangle must be even. Thus, by marking the edges of the kagome
lattice corresponding to flipped arrows, we obtain a pattern of closed
loops on the hexagonal lattice, which we denote by $L_D$; vice versa,
every loop pattern $L$ has a dimer configuration $D_L$ associated with it.
The construction is illustrated in Fig.~\ref{fig:looppatterns}a.
Note that the mapping between loop patterns and dimer configurations is
only defined relative to the reference configuration, changing to a
different reference configuration $R'$ corresponds to flipping the link
variables of the loop model according to the loop pattern $L_{R'}$.

The mapping between dimer configurations and loop patterns can be extended
to quantum states: The dimer model $\sum_D\ket{D}$ is then mapped to the
state $\sum_L\ket{L}$, a uniform superposition of all loop patterns on the
hexagonal lattice; here, the state $\ket{L}$ consists of two-level systems
which live on the edges, where the state $\ket{1}$ ($\ket{0}$) marks the
presence (absence) of a string. The latter is the ground state of the
seminal Toric Code model by Kitaev, which has $\mathbb Z_2$ topological
order.~\cite{kitaev:toriccode} Given a suitable local encoding of
$\ket{D}$, this mapping corresponds to a constant-depth unitary local
circuit, and thus, the kagome dimer model is in the same phase as the
toric code.

We can now use this mapping to construct resonating valence bond states
and dimer models with semionic statistics. To this end, consider the state
\[
\ket{\psi_{\mathrm{sem}}} = \sum_L (-1)^{n(L)}\ket{L}\ ,
\]
where $n(L)$ is the number of closed loops in $L$. It describes the ground
state of the double semion model, which also has $\mathbb Z_2$ topological
order, but differs from the Toric Code model in the statistics of its
excitations.~\cite{freedman:semionmodel} We are now ready to define the
wavefunction of the
\emph{semionic dimer model}~\footnote{ There are two reasons to define
semionic dimer models w.r.t.\ loop patterns on the honeycomb rather than
on the kagome lattice.  First, while overlap graphs of dimer configurations
yield loop configurations also on the kagome lattice, this mapping is not
surjective (i.e., relative to a given reference configuration, not every
loop pattern corresponds to a dimer pattern), and second, counting the
number of loops is ambiguous on non-trivalent lattices.
}
\begin{equation}
\left| { \psi  }_\mathrm{sem\textrm{-}dimer} \right> =
    \sum _{ D }^{  }{ { \left( -1 \right)  }^{ n({ L }_{ D }) }\left| D
\right>  } \ .
\end{equation}
Again, given a suitable encoding of $\ket{D}$, it is locally equivalent
to $\ket{\psi_\mathrm{sem}}$ and thus in the same phase as the double
semion model.  In the same way, we also define the \emph{semionic RVB
state} as
\begin{equation}
\left| { \psi  }_{ \mathrm{semRVB} } \right> =\sum _{D}^{  }
{ { \left( -1 \right)  }^{ n(L_D) }\left| \sigma
\left( D \right)  \right>  } \ .
\end{equation}

Let us point out that fixing a reference configuration necessarily breaks
the symmetry of the kagome lattice.  As we will see, this gives rise to symmetry
breaking in the semionic dimer and RVB model, while this is not the case
for the conventional dimer/RVB state. This can be understood from the fact
that changing the reference configuration corresponds to flipping a
certain loop pattern: While this is a $1$-to-$1$ mapping on the set of all
loop configurations and thus does not affect the conventional dimer and
RVB model, it changes the number of loops and thus the phases in the
semionic wavefunctions.

\section{PEPS representation
\label{sec:peps-formalism}}

In this section, we show how semionic RVB states can
be expressed using Tensor Networks. 

\subsection{Tensor Network for the semionic RVB state}

Let us start by reviewing the tensor network representations of the Toric
Code~\cite{verstraete:comp-power-of-peps,schuch:peps-sym} and of the
double semion model.~\cite{gu:stringnet-peps} Both models are superpositions of closed
loops on the hexagonal lattice, though with different sign patterns. Any
loop configuration has a dual representation in terms of Ising variables
(i.e., two-level systems) on the plaquettes, where we put a loop whenever
the plaquette variable changes, as illustrated in
Fig.~\ref{fig:looppatterns}b; we will refer to these as ``color''
variables in the following. Note that the mapping from the color
representation to loops is $2$-to-$1$, since flipping all colors yields
the same loop pattern.

\begin{figure}
\includegraphics[width=0.7\columnwidth]{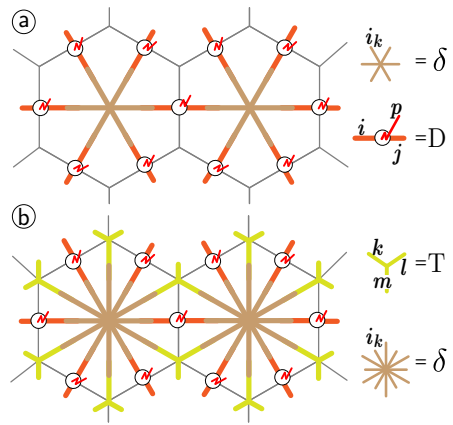}
\caption{\label{fig:tn-tcode-dsem}
Tensor network representation for \textbf{(a)} the Toric Code and
\textbf{(b)} the double semion model. 
}
\end{figure}

In terms of these color variables, a tensor network representation of the
Toric Code state can be obtained as follows: First, associate a $6$-index
Kronecker delta tensor $\delta_{i_1,\dots,i_6}$ ($i_k=0,1$,
$\delta_{i_1,\dots,i_6}=1$ iff all $i_k$ are equal) to each plaquette; it
carries the plaquette color.  Second, associate a tensor 
\begin{equation}
\label{eq:D-tensor}
D_{i,j}^p =\delta_{i\oplus j,p}
\end{equation}
which is $1$ iff $i\oplus j=p$ to each edge (where $\oplus$ denotes addition
modulo $2$). We now contract the
$i$ and $j$ index of $D$ with the indices of the $\delta$ tensors of the
adjacent plaquettes, while $p$ describes the physical index and remains
uncontracted, as indicated in Fig.~\ref{fig:tn-tcode-dsem}a. Since by
Eq.~\eqref{eq:D-tensor} $p$ is the difference of the adjacent plaquette
color variables, this exactly yields the desired sum over all loop
patterns.

In order to obtain a tensor network description of the double semion
model, we additionally need to weigh each loop with $-1$. Following
Ref.~\onlinecite{gu:stringnet-peps}, this is achieved by integrating the
curvature of each loop by assigning a phase of $+i$ ($-i$) to each vertex
with exactly one (two) adjacent black plaquettes, giving a phase of $(\pm
i)^6=-1$ for each closed loop. In terms of tensor networks, this is
achieved by changing the $\delta$ tensor inside the plaquettes to a 
$12$-index $\delta$ tensor, where the additional indices are contracted
with tensors 
\[
T_{klm} = \left\{\begin{array}{r@{\quad}l}
    +i & \mbox{if \ }k+l+m= 1\\
    -i & \mbox{if \ }k+l+m= 2\\
    1 & \mbox{else}
\end{array}\right.
\]
placed on the vertices, as shown in Fig.~\ref{fig:tn-tcode-dsem}b.

\begin{figure}
\includegraphics[width=0.7\columnwidth]{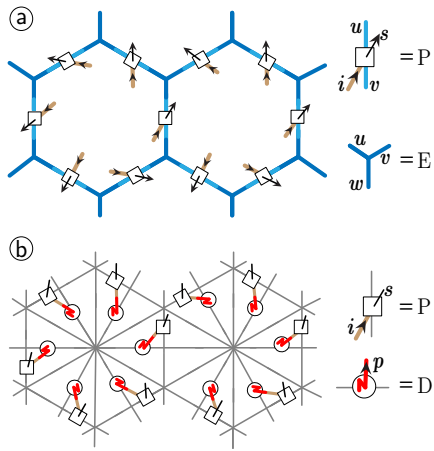}
\caption{\label{fig:tn-semRVB}
Tensor network representation of the semionic RVB state. The tensor
network in panel (a) needs to be combined with the tensor network
for the double semion model (Fig.~\ref{fig:tn-tcode-dsem}b), yielding a
tensor network with the structure indicated in panel (b).
}
\end{figure}

We are now in the position to construct the PEPS representation for the
conventional and semionic RVB state. To this end, we start from the tensor
network representation of the Toric Code or double semion model,
respectively, and place a tripartite tensor $E_{uvw}$ ($u,v,w=0,1,2$), at
each vertex of the honeycomb lattice, with one degree of freedom
associated to each edge (i.e., the vertices of the kagome lattice).  $E$
is of the form
\begin{equation}
\label{eq:E-tensor-def}
E_{uvw} = \left\{\begin{array}{c@{\quad}l}
    1 & u=v=w=2 \\
    \varepsilon_{uvw} & \mbox{otherwise}
\end{array}\right.
\end{equation}
with $\varepsilon_{uvw}$ the fully antisymmetric tensor. $E$ thus
describes a triangle with either one singlet in the $\{0,1\}$ subspace or
with no singlet at all, where the absence of a singlet is indicated by a
$2$. Note that depending on the chosen orientation of the singlets,
the signs in $\varepsilon_{uvw}$ might have to be modified.  At each site, we then add
a tensor $P_{i;u,v}^s$ ($i=0,1$, $u,v=0,1,2$, $s=0,1$) which picks a
singlet from the $E$ tensor of one of the adjacent triangles (and enforces
the index of the other to be $2$), as prescribed by the value of the loop
and the reference configuration; this is,
\[
P_{i;u,v}^s = \left\{\begin{array}{r@{\quad}l}
    1 & s=u,\,v=2,\,i=0\\
    1 & s=v,\,u=2,\,i=1\\
    0 & \mbox{otherwise}
\end{array}\right.
\]
where $i$ is contracted with the ``physical'' index of the underlying loop
model, $s$ is the physical index of the RVB state, and $u$ and $v$ are
contracted with the indices of the adjacent $E$'s as prescribed by the
reference configuration; the construction is illustrated in
Fig.~\ref{fig:tn-semRVB}.  Note that the tensor network can be simplified
by grouping the tensors $D$ and $P$ into a single tensor.

Along the same line, we can also construct a tensor network
representation of normal or semionic dimer
models.~\cite{schuch:rvb-kagome} To this end, we replace $P_{i;u,v}^s$ by
$(P_\perp)_{i;u,v}^{s,t} = P_{i,u,v}^{s}\delta_{i,t}$ ($t=0,1$) with
physical indices $s$ and $t$; this is, the loop degree of freedom $i$
remains physical and serves as an ``indicator qubit'' which allows to
locally distinguish different singlet configurations; it is easy to see
that the resulting dimer model is locally unitarily equivalent to the
corresponding loop model.~\footnote{This can be seen by adding
a spin-$\tfrac12$ state at every edge of the kagome lattice, and applying
a unitary first on the up and then on the down triangles which creates
singlets depending on the arrow pattern corresponding to the loop
configuration.}
Following Ref.~\onlinecite{schuch:rvb-kagome}, we can generalize this to
construct an interpolation from RVB states to dimer models, by using
\begin{equation}
\label{eq:P-theta-def}
P(\theta)^{s,t}_{i,u,v}= P_{i,u,v}^{s}w(\theta)_{i,t}\ ,
\end{equation}
where $w(\theta)_{i,t} = 1+(-1)^{i+t}\theta$; here, $\theta=1$ corresponds
to the dimer model and $\theta=0$ to the RVB state (with the $t$ qubits in
the $\ket+$ state).

\subsection{Ground state manifold
\label{sec:tn:gs-manifold}}

In the preceding subsection, we have shown how to write the semionic RVB
state as a PEPS.  Yet, the double semion model is a topological model with
a $4$-fold degenerate ground space. In the following, we will show how to
parametrize the ground space in terms of its PEPS representation.

\begin{figure}
\includegraphics[width=0.95\columnwidth]{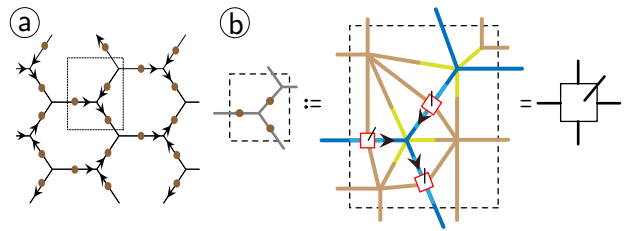}
\caption{\label{fig:square-blocking}
\textbf{(a)} Blocking of the hexagonal and kagome lattice into square
blocks.  \textbf{(b)} Structure of the blocked tensor for the double
semion model and semionic RVB state. For the double semion model, we can
introduce the blue ``string'' indices which are redundant but ease the
transition to the semionic RVB state.
}
\end{figure}

\begin{figure}[t]
\includegraphics[width=0.75\columnwidth]{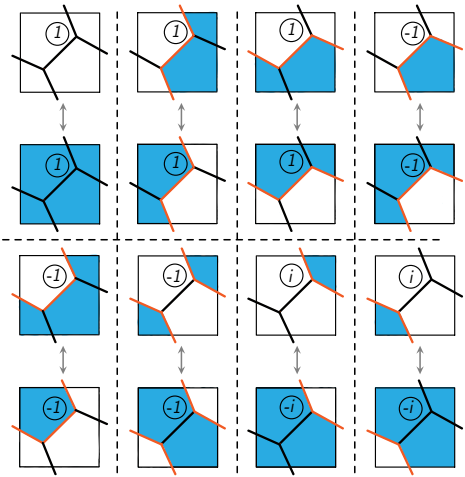}
\caption{\label{fig:16-confs}
Possible configurations of blocked tensor of the double semion model
with the corresponding phases. The configurations form pairs (on top of
each other) which are related by flipping the color, and thus have the
same physical configuration, giving rise to a virtual $\mathbb Z_4$
symmetry (see text).
}
\end{figure}

\begin{figure}[t]
\includegraphics[width=0.7\columnwidth]{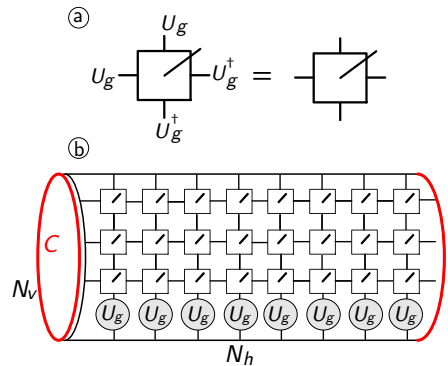}
\caption{\label{fig:Ug-closure}
\textbf{(a)} Virtual symmetry ($G$-invariance) of the PEPS tensor.
\textbf{(b)} Parametrization of the ground state manifold in terms of the
symmetry: The minimally entangled states are obtained by placing a
``flux'' string of group elements $U_g$ along the horizontal closure, and
projecting the boundary conditions onto irreducible representations
(``charges'') of $U_g^{\otimes N_v}$.
}
\end{figure}

We start by considering the PEPS representation of the double semion
model; the generalization to the semionic RVB will be immediate.  We
first rewrite its PEPS representation by blocking three sites into a unit cell,
thereby obtaining a square lattice, as indicated in
Fig.~\ref{fig:square-blocking}a. (While
the blocking is not required for the ground space parametrization, it will
be useful for the numerical implementation.) For the blocked tensor, the
$\delta$ tensor in each plaquette is decomposed as a contraction of four
$\delta$ tensors, one for each block; overall, this results in a
blocked tensor which has two indices at each side, one for each plaquette
color (indicated by brown lines in Fig.~\ref{fig:square-blocking}b). In
order to make the transition to the semionic RVB state easier, we add a
third (redundant) index at each side, which is equal to the corresponding
loop variable, i.e., the difference of the adjacent plaquette variables
(the blue line in Fig.~\ref{fig:square-blocking}b; this index will later
be replaced by the ``singlet or $2$'' index of $E$).  The auxiliary degrees
of freedom of this new tensor have $16$ non-zero values, which are
illustrated in Fig.~\ref{fig:16-confs} with the corresponding amplitudes. 
These
$16$ configurations come in pairs which are related by flipping all
plaquette colors, and thus correspond to the same physical state. It is
now easy to see that this gives rise to a virtual symmetry of the tensor
which corresponds to flipping all plaquette colors and in addition
adjusting the phases depending on the loop configuration.  This is, the
tensor is invariant under a virtual symmetry action $U_g$ on all four
sides  as shown in Fig.~\ref{fig:Ug-closure}a. Here, $U_g = \zeta^g$ is a
$\mathbb Z_4$ symmetry action, $g=\{0,1,2,3\}$, with 
\begin{equation}
\label{eq:zeta-def}
\zeta = X\otimes \eta \otimes X\ ,
\end{equation}
where $X = \left(\begin{smallmatrix}0&1\\1&0\end{smallmatrix}\right)$ acts
on the color indices and
$\eta = \left(\begin{smallmatrix}1&0\\0&i\end{smallmatrix}\right)$ acts on
the loop index.

The invariance of the PEPS tensor under this group action is closely
related to topological order, and it can be used to parametrize different
ground states:~\cite{schuch:peps-sym} In particular, on a cylinder we 
can construct topologically distinct ground states by placing strings
of $U_g$ along the cylinder axis and projecting the left/right boundary
condition of the cylinder onto irreducible representations of the group
action, as shown in Fig.~\ref{fig:Ug-closure}b. (Intuitively, this can be understood
from the fact that using the symmetry of the tensor these strings can be
freely moved through the lattice and thus should not affect the state
locally.) This
way, we obtain $16$ possible states, which are labelled by a group
element (flux) $\phi=0,1,2,3$ and an irreducible representation (charge)
$c = 1,i,-1,-i$ of $\mathbb Z_4$. Yet, the double semion model has only
four distinct ground states, and there is indeed a redundancy in this
description: As is discussed in Appendix~\ref{sec:app:groundstates} and
as can also be checked numerically, eight
of these states have norm zero, while the remaining eight form pairs which
describe the same ground state; we thus find that the four ground states
of the double semion model correspond to the following flux and charge
labels:
\begin{equation}
\label{eq:topo-groundstates}
\begin{array}{ll@{\mbox{\ or\ }}l}
\ket{\psi_1} & \ \leftrightarrow\ (\phi=0,c=1)&(\phi=2,c=-1)
\\
\ket{\psi_s} & \ \leftrightarrow\ (\phi=1,c=i)&(\phi=3,c=-i)
\\
\ket{\psi_{\bar{s}}} & \ \leftrightarrow\ (\phi=1,c=-i)&(\phi=3,c=i)
\\
\ket{\psi_b} & \ \leftrightarrow\ (\phi=0,c=-1)&(\phi=2,c=1)
\end{array}
\end{equation}
Here, we have labelled the ground states by the particle types of the
model (the trivial particle, a conjugate pair of semions, and a boson);
the identification can be e.g.\ understood by noting that the semionic
ground states correspond to eigenstates of a loop operator of a bound
state of an electric and a magnetic particle with charge/flux $\tfrac14$
in a $\mathbb Z_4$ double model,~\cite{kitaev:toriccode} which has
semionic statistics.

In the case of semionic RVB states, the parametrization of the ground
space is exactly analogous, with the only difference that $\eta$ has to be
replaced by $\eta = \mathrm{diag}(1,1,i)$ or $\eta=\mathrm{diag}(i,i,1)$,
depending on whether $\{0,1\}$ or $2$ corresponds to the presence of a
string in the double semion model, as determined by the reference
configuration.

\section{\label{sec:results}Results}

In this section, we present a range of numerical results on semionic RVB
states and extensions thereof, obtained using exact PEPS contraction
techniques (see Appendix~\ref{sec:app:numerical-optimization} for
details).

\begin{figure}[t]
\includegraphics[width=0.75\columnwidth]{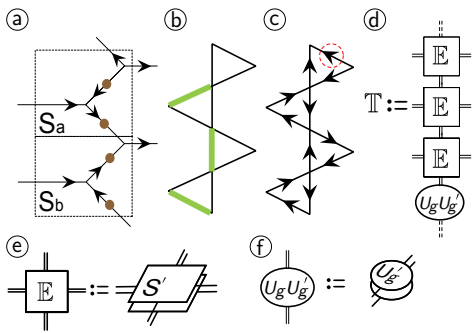}
\caption{\label{fig:block-refconf-top}
\textbf{(a)} $6$-site unit cell used for the numerical simulations.
\textbf{(b)} Reference configuration used. \textbf{(c)} Energetically
optimal singlet orientation for the semionic RVB. \textbf{(d-f)}
Construction of the transfer operator (d), with the form of the individual
ket-bra tensors and closures shown in (e) and (f),
with $S'$ the blocked tensor of panel (a).
}
\end{figure}

\subsection{Prerequisites}

First, we need to fix a reference configuration of dimers. To this end, we
choose a $6$-site unit cell as indicated in
Fig.~\ref{fig:block-refconf-top}a; this is the smallest unit cell for
which reference configurations exist which allow to tile the lattice.  We
use the reference configuration shown in
Fig.~\ref{fig:block-refconf-top}b; all other reference configurations are
related to this configuration by a symmetry transformation of the lattice,
cf.~Appendix~\ref{sec:app:dep-on-reference}.

We focus on the study of PEPS on infinite cylinders, see
Fig.~\ref{fig:Ug-closure}b, where we extrapolate in the cylinder
circumference $N_v$. A central object in this study is the so-called
transfer operator, which is obtained by contracting the ket and bra layer
of the PEPS and considering one column, as shown in
Fig.~\ref{fig:block-refconf-top}d-f.
Such a transfer operator can be constructed for the overlap
of two arbitrary ground states $\ket{\psi_p}$ and $\ket{\psi_q}$, and we
denote it by $\mathbb T_p^q$. The largest eigenvalue $\gamma_p^q$ of
$\mathbb T_p^q$ allows to determine the overlap
$\langle\psi_q\ket{\psi_p}/(\big\vert\ket{\psi_p}\big\vert\,
\big\vert\ket{\psi_q}\big\vert) \sim
(\hat\gamma_p^q)^{N_h}$ with
$\hat\gamma_p^q=\gamma_p^q/(\gamma_p^p\gamma_q^q)^{1/2}$; in particular,
if $\hat\gamma_p^q\rightarrow\mathrm{const.}<1$ as $N_v\rightarrow\infty$,
$\ket{\psi_p}$ and $\ket{\psi_q}$ are orthogonal, whereas if
$\hat\gamma_p^q = 1+o(1/N_v)$, $\ket{\psi_p}$ and $\ket{\psi_q}$ describe
the same state in the thermodynamic limit.~\cite{schuch:topo-top} Note
also that $-1/\log \hat\gamma_p^q$ relates to the ``size'' of the topological
non-trivial excitations which couple ground states $p$ and $q$.  Also of
interest are the second largest eigenvalues $\tau_p^q$; in particular,
$\xi_p=-1/\log(\tau_p^p/\gamma_p^p)$ bounds the correlation length in the
ground state $\ket{\psi_p}$.

\begin{figure}
\includegraphics[width=\columnwidth]{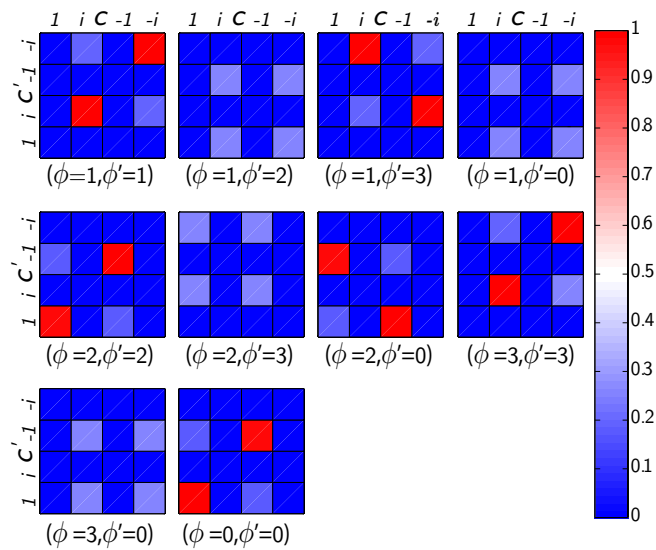}
\caption{\label{fig:cauchyschwarz} 
Absolute value of the largest eigenvalues $\gamma_p^q$ of the transfer
operator $\mathbb{T}$, normalized to $1$.  $\phi$ ($\phi'$) and $c$ ($c'$)
specify flux and charge in the ket (bra) layer of $\mathbb{T}$, i.e.,
$p\equiv (\phi,c)$ and $q\equiv (\phi',c')$. One can easily verify that
only the four distinct states of Eq.~(\ref{eq:topo-groundstates}) remain.
Note that $(\gamma_p^q)=(\gamma_q^p)^*$.
}
\end{figure}

\subsection{Results for the semionic RVB state}

In the following, we present our numerical results for the semionic
RVB state. All results have been obtained with the reference configuration
and with the singlet orientation shown in
Fig.~\ref{fig:block-refconf-top}c; we have
found this singlet orientation to be the (non-unique) one which minimizes
the energy of the semionic RVB state as an ansatz for the kagome
Heisenberg antiferromagnet.

\subsubsection{Overlap of all sectors}

To start with, we have considered the parametrization of the ground space
in terms of fluxes and charges, as explained in
Sec.~\ref{sec:tn:gs-manifold}. This
parametrization gives a total of $16$ states (corresponding to $\phi=0,1,2,3$
and $c=1,i,-1,-i$), for which we show $\gamma_p^q$ (normalized by
$\max\gamma_p^q$) for all possible $16$ values of $p$ and $q$ in
Fig.~\ref{fig:cauchyschwarz}.
One can clearly see that eight of these states have zero norm, while the
remaining eight form pairs for which $\gamma_p^q =
(\gamma_p^p\gamma_q^q)^{1/2}$, thereby yielding the four ground states given in
Eq.~(\ref{eq:topo-groundstates}). On the other hand, for the remaining four states,
$\gamma_p^q< (\gamma_p^p\gamma_q^q)^{1/2}$; this is, they indeed describe
four distinct ground states of the resulting model.

\subsubsection{Interpolation, phase, and correlations}

\begin{figure}
\includegraphics[width=\columnwidth]{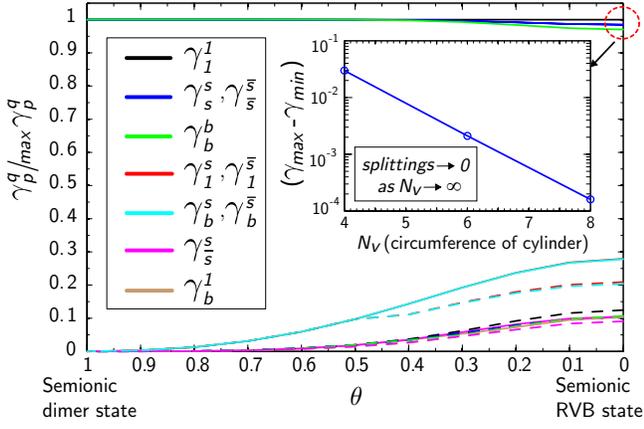}
\caption{
\label{fig:interpol1}
Largest (solid lines) and second largest (dashed lines) eigenvalues for
the different non-zero blocks of the transfer operator $\mathbb T_p^q$
along the interpolation from the semionic dimer state to the semionic RVB
state for $N_v=4$ (YC8 in the notation of
Ref.~\onlinecite{yan:heisenberg-kagome}). All $\gamma_p^p\rightarrow1$ exponentially as
$N_v\rightarrow\infty$ (inset),  while the other eigenvalues stay bounded
away, showing that the different topological ground states remain
orthogonal and of equal norm, and the correlation length remains bounded,
showing no sign of a phase transition
(with $\xi\approx0.42$ for the semionic RVB state), demonstrating
that the semionic RVB state is in the phase of the double semion model. 
}
\end{figure}

Next, we have studied whether the semionic RVB state is in the same phase
as the semionic dimer model (and thus the double semion model). To this
end, we have used the interpolation $P(\theta)$ between the semionic RVB
and the semionic dimer model given in Eq.~(\ref{eq:P-theta-def}). The
results are shown in Fig.~\ref{fig:interpol1}: We find that the largest
eigenvalues $\gamma_p^p$ for all four ground states remains equal (up to an
splitting exponentially small in $N_v$, see inset), while $\gamma_p^q$ for $p\ne q$
remains strictly smaller, which shows that the four ground states remain
stable and orthogonal to each other.  At the same time, the second largest
eigenvalues remain bounded, ruling out a diverging correlation length.
Together, this provides compelling evidence that the semionic RVB state is in
the phase of the double semion model.  From this data, we can also extract
the correlation length at the semionic RVB point, which we find to be
$\xi\approx 0.42$, and the coherence length of topologically
non-trivial semionic and bosonic excitations, $\xi_{\mathrm{sem}} \approx
0.81$ and $\xi_{\mathrm{bos}} \approx 0.44$.
(For comparison, the values obtained for the
RVB transfer operator~\cite{schuch:topo-top} are $\xi\approx 0.79$,
$\xi_\mathrm{spinon}\approx 1.21$, and $\xi_\mathrm{vison}\approx0.79$.)

\subsubsection{Ansatz for the kagome Heisenberg model}

\begin{figure}
\includegraphics[width=\columnwidth]{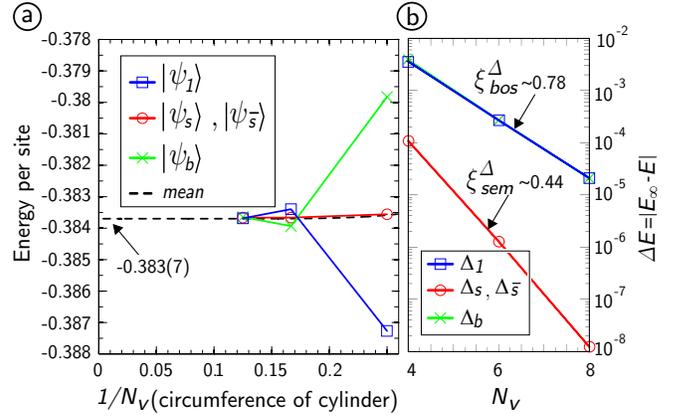}
\caption{
\label{fig:optimalnrgypersite}
\textbf{(a)}
Energy per site for the semionic RVB state w.r.t.\ the Heisenberg
Hamiltonian as a function of $1/N_v$ for the different topological
sectors and $N_v=4,6,8$; the extrapolated energy per site is
$E_\infty=-0.383(7)$.  \textbf{(b)} Exponentially vanishing finite-size
difference between the energy per site and the extrapolated value for
$N_v\rightarrow\infty$, with corresponding length scales
$\xi^\Delta_{\mathrm{bos,sem}}$.
}
\end{figure}

\begin{figure}[t]
\includegraphics[width=\columnwidth]{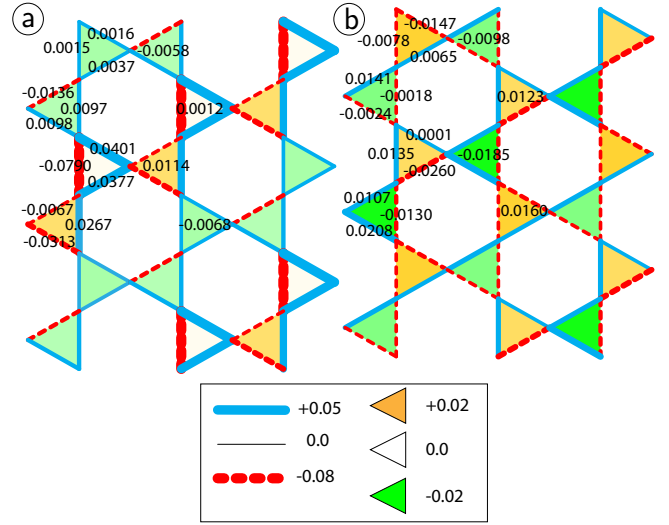}
\caption{
\label{fig:simplexenergydev}
Variations in the energy per link and the energy per triangle due to the
explicit symmetry breaking induced by the choice of the reference
configuration, for (a) the semionic RVB state with $N_v=8$ and (b) the
semionic simplex RVB state with $N_v=6$.  The thickness and color of each
line indicates the magnitude and sign of the difference $\langle \bm S_i\cdot
\bm S_j\rangle - E_\mathrm{semRVB}^{\mathrm{(simplex)}}/2$; the intensity
and color of each triangles indicates the deviation of the sum of bond
energies from $\tfrac32 E_\mathrm{semRVB}^{\mathrm{(simplex)}}$. The
energies for the different topological sectors converge exponentially and
break the symmetry in an identical way.
}
\end{figure}

A main motivation for studying semionic RVB states was that they might form an
alternative to the conventional RVB state as an ansatz for the ground
state of the kagome Heisenberg antiferromagnet.  We have thus computed the
energy of the kagome AFM for all four ground states sectors and for
different values of $N_v$; the results are shown in
Fig.~\ref{fig:optimalnrgypersite}.  We find that the energies of the
different ground states converge exponentially in $N_v$, as is expected
for topologically degenerate ground states, with an extrapolated energy
per site in the thermodynamic limit of $E_{\mathrm{semRVB}} = -0.383(7)$.
Fixing a reference configuration explicitly breaks translational
invariance, leading to different energies for the different edges in a
unit cell; the extent of translational symmetry breaking is shown in
Fig.~\ref{fig:simplexenergydev}a, it is (approximately) identical for all
topological sectors.

\subsection{\label{sec:subsec:simplex-rvb}
Semionic simplex RVB states}

The energy $E_\mathrm{semRVB} = -0.383(7)$ of the semionic RVB state as an
ansatz for the Heisenberg antiferromagnet is higher than the energy found
for the conventional RVB state,~\cite{poilblanc:rvb-boundaries}
$E_\mathrm{RVB}=-0.3931$, which might suggest that the conventional RVB
state is a more accurate description of the ground state of the kagome
HAFM. On the other hand, it has been found in
Ref.~\onlinecite{poilblanc:simplex-rvb} that the energy of the RVB can be
significantly improved by introducing only two extra parameters, an ansatz
termed simplex RVB state. In the following, we study the same
two-parameter family of semionic simplex RVB states, and find that by
tuning these parameters, the family of semionic simplex RVB states
achieves a variational energy even slightly below that of simplex RVB
states.

\subsubsection{Semionic simplex RVB ansatz}

Let us first briefly sketch the idea of (semionic) simplex RVB
states; for details, we refer the reader to
Ref.~\onlinecite{poilblanc:simplex-rvb}.  The construction is based on the observation
that defect triangles, i.e., those without singlets, are energetically
unfavorable. Zeng and Elser~\cite{zeng:kagome-nnn-ham} showed that the
energy can be significantly improved by allowing for singlets between
next-nearest neighbors which are obtained by increasing the weight of the
spin-$\tfrac12$ subspace on each triangle. In the simplex RVB
construction, a similar idea is used
to improve the energy without increasing the bond dimension of the
blocked tensors: First, the number of right-pointing defect triangles is
reduced by changing $E_{ijk}$, Eq.~\eqref{eq:E-tensor-def}, on the
right-pointing triangles to $E'_{ijk}$ with $E'_{222}=\beta<1$ and
$E'_{ijk}=E_{ijk}$ otherwise, and second, the energy of the left-pointing
triangles is improved by multiplying the spins on each of those triangles
by $\openone - \alpha \mathbb P_{3/2}$, with $\mathbb P_{3/2}$ the
projector onto the spin $3/2$ (i.e., permutationally invariant) subspace;
since the right-pointing triangles sit inside the tensor, this does not
increase the bond dimension $D$ of the PEPS.

\subsubsection{Ansatz for the kagome HAFM}

\begin{figure}
\includegraphics[width=0.9\columnwidth]{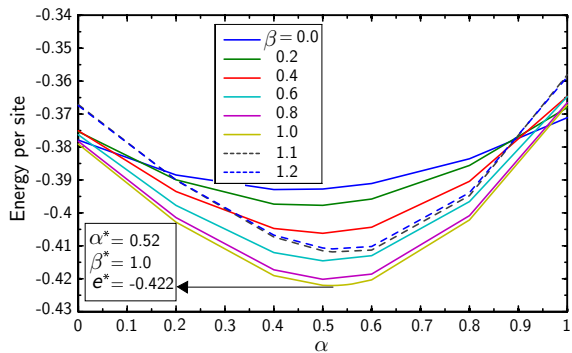}
\caption{\label{fig:simplex-optimization}
Optimization for the simplex RVB: Energy per site as a function of the two
parameters $\alpha$ and $\beta$ for $N_v=4$ (see text).
}
\end{figure}

\begin{figure}
\includegraphics[width=\columnwidth]{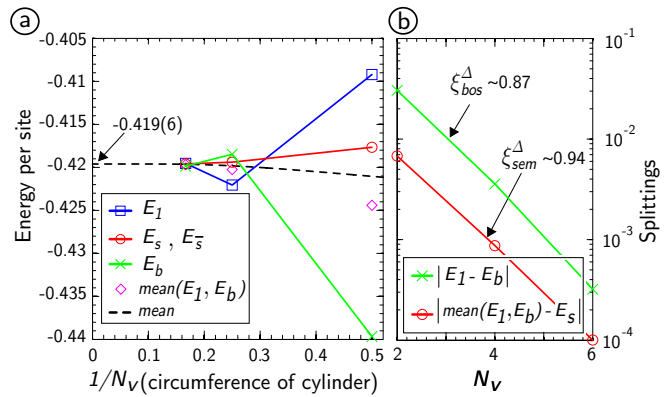}
\caption{
\label{fig:simplexenergies}
Energy per site for the semionic simplex RVB state w.r.t.\ the Heisenberg
Hamiltonian as a function of $1/N_v$ for the different topological
sectors; the extrapolated energy per site is $E_\infty=-0.419(6)$.
\textbf{(b)} Exponentially vanishing finite-size splitting
for $N_v\rightarrow\infty$, with length scales
$\xi^\Delta_{\mathrm{bos,sem}}$.
}
\end{figure}

Fig.~\ref{fig:simplex-optimization} shows the energy for the HAFM
Hamiltonian as a function of $\alpha$ and $\beta$.  We find that for
$N_v=4$ the minimum is obtained for a value of $\alpha\approx 0.52$ and
$\beta=1$, and with the canonical singlet orientation of
Fig.~\ref{fig:kagomes}a, and we fix these values for the extrapolation in
$N_v$.  Fig.~\ref{fig:simplexenergies}a shows the energy per site as a
function of $1/N_v$ for the different topological sectors. The energies
again converge exponentially in $N_v$ (Fig.~\ref{fig:simplexenergies}b),
and we find the
extrapolated energy to be
$E^{\mathrm{simplex}}_{\mathrm{semRVB}}=-0.4196$, which is a significant
improvement over the energy $E_{\mathrm{semRVB}}$ of the semionic RVB
state, and in fact even slightly below the energy found for the
conventional simplex RVB,~\cite{poilblanc:simplex-rvb}
$E^{\mathrm{simplex}}_{\mathrm{RVB}} = -0.4181$.

Again, due to the choice of a reference pattern the system breaks the
symmetries of the lattice.  The variation in energy over the different
bonds is shown in Fig.~\ref{fig:simplexenergydev}b, and we find that it is
significantly reduced as compared to the unoptimized semionic RVB state;
in fact, the observed variations in energy are comparable to those found
in variational calculations for the kagome HAFM for certain choices of
boundary
conditions.~\cite{yan:heisenberg-kagome,depenbrock:kagome-heisenberg-dmrg}

\begin{figure}[t]
\includegraphics[width=0.9\columnwidth]{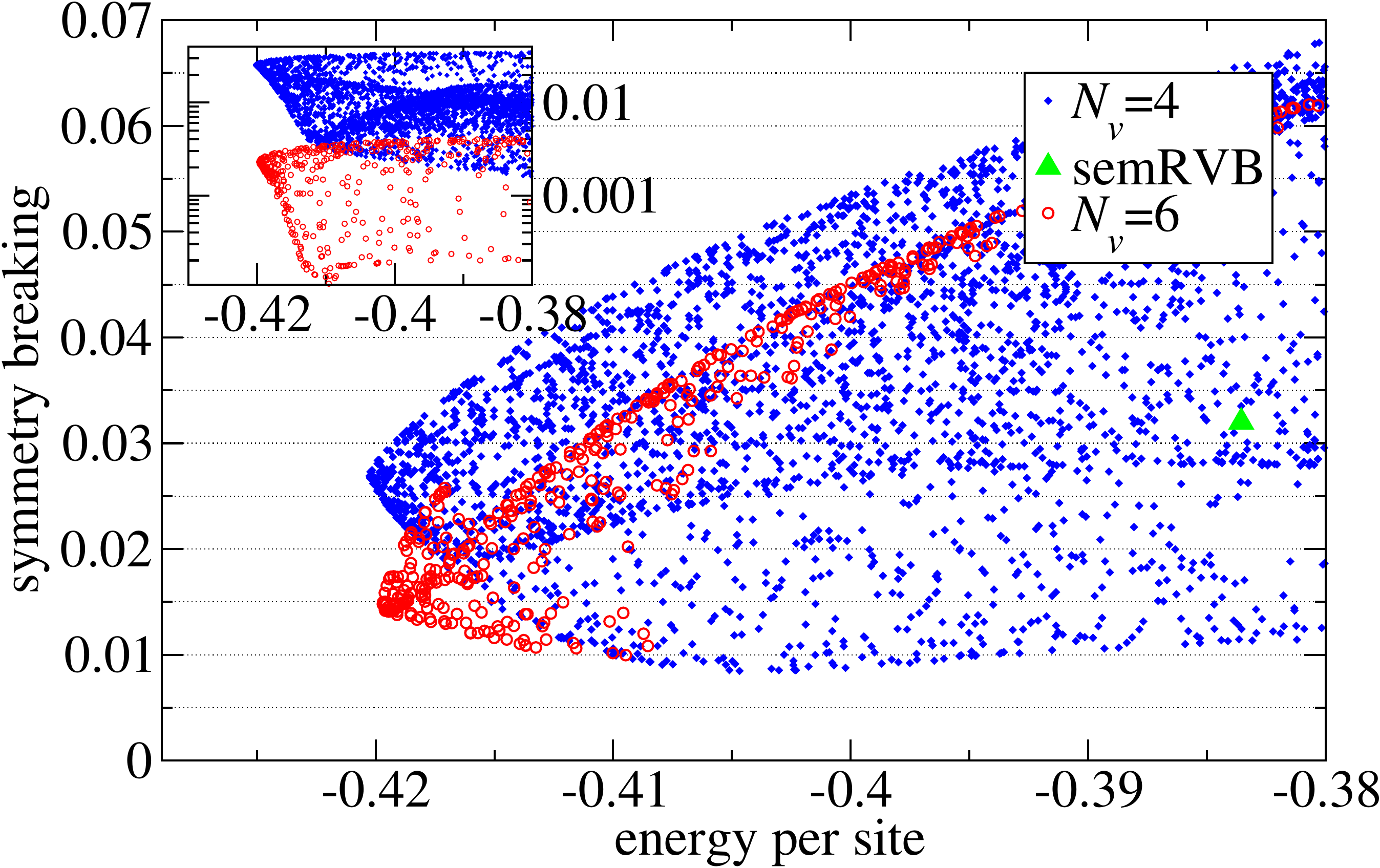}
\caption{\label{fig:variations-vs-energy}
Relation between symmetry breaking and variational energy for the semionic
simplex RVB state for $N_v=4$ and $N_v=6$ (for randomly chosen values of
$\alpha$ and $\beta$ in a neighborhood of the optimum). The $x$ axis
shows the energy per site, and the $y$ axis gives the standard deviation
of the $12$ inequivalent energies in the unit cell (averaged over
$\ket{\psi_1}$ and $\ket{\psi_b})$. To be able to identify finite-size
effects, we plot in the inset the standard deviation of the difference
between the energy of $\ket{\psi_1}$ and $\ket{\psi_b}$ for each link,
which measures the finite-size effects, vs.\ the energy. We find that for
$N_v=6$, the symmetry breaking is clearly above the finite size effects,
suggesting it is favorable to break the symmetry to reach the optimal
energy.  The green triangle corresponds to the semionic RVB state.
}
\end{figure}

To better understand the nature of the symmetry breaking, we have also
investigated the dependence between the variational energy and the degree
of symmetry breaking within the two-parameter family of semionic simplex
RVB. The results are shown in Fig.~\ref{fig:variations-vs-energy} for
$N_v=4$ and $N_v=6$, where we plot the standard deviation
of the energy over the different links (averaged over $\ket{\psi_1}$ and
$\ket{\psi_b}$ for each link) vs.\ the variational energy. As one can see,
there are strong finite-size effects. To be able to nevertheless single
out the effects related to symmetry breaking in the thermodynamic limit, we
additionally plot in the inset the difference between the energy per link for
$\ket{\psi_1}$ and for $\ket{\psi_b}$ (averaged using the standard deviation)
vs.\ the energy, which quantifies the finite size effects. We find that
while for $N_v=4$, the measured symmetry breaking is comparable to the
finite-size effects, for $N_v=6$ the finite size effects are significantly
lower than observed symmetry breaking, which shows that the symmetry
breaking will survive in the thermodynamic limit.  We also find that the
symmetry breaking at the energetic minimum is clearly larger than the
minimal symmetry breaking compatible with the ansatz, which suggests that
double-semion topological order in the kagome HAFM might favor symmetry
breaking, though finite size effects cannot entirely be ruled out.
(This should be contrasted with the behavior of the conventional simplex
RVB, which we show in Appendix~\ref{sec:app:symbreaking-normal-rvb}.)
Interestingly, we also found that the average energies in the left and
right triangles are identical, despite the lattice symmetry breaking.

\subsubsection{Interpolation, phase, and correlation length}

\begin{figure}[t]
\includegraphics[width=\columnwidth]{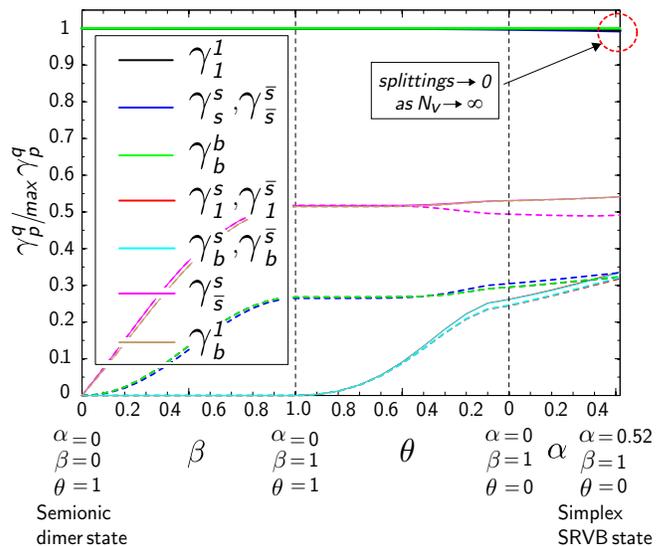}
\caption{\label{fig:interpolsimplex}
Largest (solid lines) and second largest (dashed lines) eigenvalues for
the different non-zero blocks of the transfer operator $\mathbb T_p^q$
along the interpolation from the semionic dimer state to the semionic
simplex RVB
state for $N_v=6$ (YC12) where the interpolation is performed in three
steps as indicated. As for the semionic RVB (Fig.~\ref{fig:interpol1}), we
find that the system stays topologically ordered with no sign of a phase
transition and finite correlation length, showing that the semionic
simplex RVB does not break spin rotation symmetry and is in the phase of
the double semion model.
}
\end{figure}

Let us now study whether the optimized semionic simplex RVB is still in
the phase of the double semion model.  This is particularly important since
the optimal value $\beta=1$ corresponds to entirely ruling out defects on
``right'' triangles, which correspondingly rules out certain string
configurations on the corresponding vertices of the honeycomb lattice.
While one might think that this rules out certain loop patterns even
on a global scale, one can easily verify that this is not the case as long
as the reference configuration is different on the two ``right'' triangles
in a unit cell (which is always the case).

In order to make sure that choosing $\beta=1$ does not give rise to a phase
transition, we therefore start by interpolating from the semionic dimer
model to the point $\beta=1$. The leading eigenvalues of the transfer
operator along this interpolation are shown in the left part of
Fig.~\ref{fig:interpolsimplex}: We find that the transfer operator remains
gapped and thus the semionic dimer model with
$\beta=1$ is in the same phase as the original semionic dimer state, but
exhibits an effective length scale of $-1/\log(0.5)$.  Both these facts can
be understood from a renormalization argument:  By blocking a $1\times 2$
unit cell given by the reference configuration and renormalizing the
allowed loop configurations, one finds that all loop patterns can be
obtained after one RG step, yet with different weights, giving rise to an
effective length scale in the system.  

We then continue to interpolate from the dimer state with $\beta=1$ to the
corresponding semionic RVB state using $\theta$,
Eq.~(\ref{eq:P-theta-def}) (middle part of
Fig.~\ref{fig:interpolsimplex}), and finally increase $\alpha$ to its
optimal value (note that the last interpolation is not defined for the dimer
state).  Along the whole interpolation, the spectrum of the transfer
operator exhibits no sign of a phase transition---the correlation length
is finite, and the different topological sectors remain orthogonal
throughout. At the semionic simplex RVB point, we obtain a correlation
length of $\xi \approx 0.90$ for topologically trivial and
$\xi_{\mathrm{sem}} \approx 0.91$ and $\xi_{\mathrm{bos}}\approx 1.63$ for
topologically non-trivial semionic and bosonic excitations, respectively;
remarkably, the ordering of $\xi_{\mathrm{bos}}$ and $\xi_{\mathrm{sem}}$
has changed as compared to the semionic RVB state.

\begin{figure*}[t]
\includegraphics[width=\textwidth]{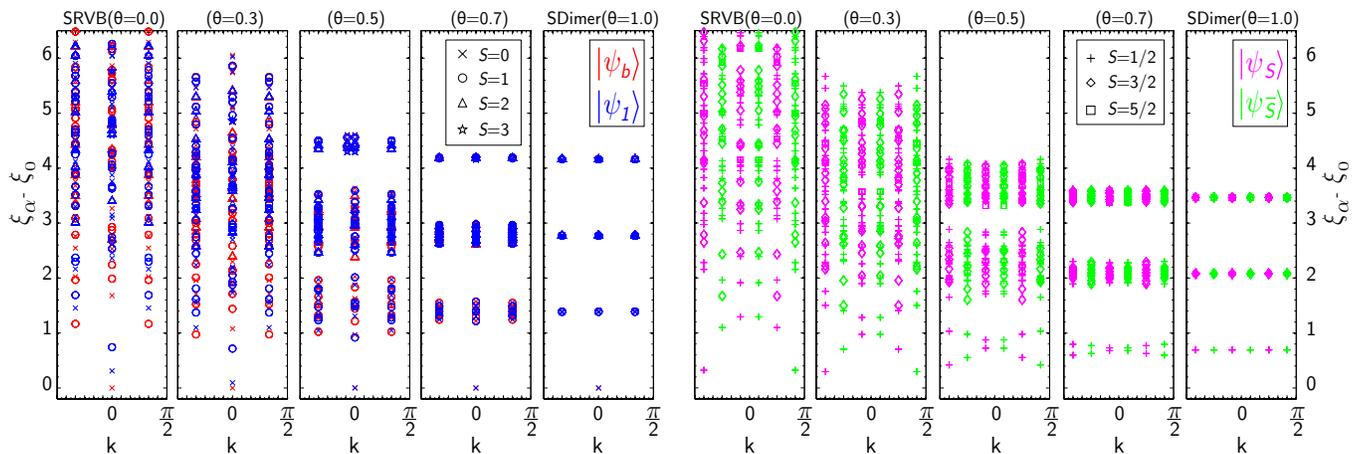}
\caption{
\label{fig:es}
Entanglement spectrum (ES) for the interpolation from the semionic dimer
state ($\theta=1$) to the semionic RVB state ($\theta=0$) for an infinite
cylinder with $N_v=6$.  The left (right) panel shows the integer
(half-integer) spin sector of the ES, which correspond to the bosonic
(semionic) ground states and are not coupled by the entanglement
Hamiltonian. The momentum is restricted to $-\tfrac\pi2\le k \le
\tfrac\pi2$ due to the blocking of two unit cells
(Fig.~\ref{fig:block-refconf-top}a).
}
\end{figure*}

\begin{figure}[b]
\includegraphics[width=0.9\columnwidth]{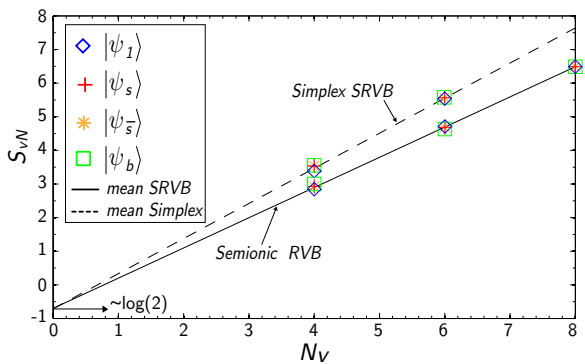}
\caption{
\label{fig:vnentropy}
Entanglement entropy $S_\mathrm{vN}$ for the semionic (solid line) and semionic simplex 
(dahsed line) RVB state, where the fit is w.r.t.\ the average of the four
topological sectors. In both cases, we find a topological correction
of $\log(2)$, indicative of a $\mathbb Z_2$ topological phase.
}
\end{figure}

\subsection{Entanglement properties}

An important way to characterize topologically ordered phases is via their
entanglement properties. The exact description of (simplex) semionic RVB
states in terms of PEPS allows to exactly compute their entanglement
properties,~\cite{cirac:peps-boundaries} which we analyze in the
following. The fits are w.r.t.\ the average over the four sectors.

\subsubsection{Entanglement entropy}

First, we have computed the (von Neumann) entanglement
entropy~\cite{levin:topological-entropy,kitaev:topological-entropy} of the
semionic RVB and semionic simplex RVB for a bipartition of the minimally
entangled states of Eq.~(\ref{eq:topo-groundstates}) into two
half-cylinders.  The results are shown in Fig.~\ref{fig:vnentropy}. By
fitting the entanglement as $S_{\mathrm{vN}} \sim c N_v -\gamma$, we find
that both states exhibit a topological
correction $\gamma \approx \log(2)$, in accordance with their $\mathbb Z_2$
topological nature.

\subsubsection{Entanglement spectrum}

We have also computed the entanglement spectrum using PEPS-based
methods.~\cite{cirac:peps-boundaries}  The results are shown in
Fig.~\ref{fig:es} separately for integer spin (bosonic ground
states) and half-integer spin (semionic ground states) excitations, with
$k$ the momentum w.r.t.\ to a $3$-site unit cell
(Fig.~\ref{fig:square-blocking}b), where the $2\times1$ blocking due to 
the reference configuration (Fig.~\ref{fig:block-refconf-top}a,b) restricts
$k$ to the reduced Brillouin zone $k=-\tfrac\pi2\dots\tfrac\pi2$; note
that for the semionic sectors, $k$ is shifted by $2\pi/(4N_v)$ due to the
presence of a $1/4$ flux in the system. (In the PEPS representation,
this is reflected in the fact that the translation operator is ``dressed''
with $U_g$, i.e., it translates and grows a string of $U_g$ at the same
time, cf.~Fig.~\ref{fig:Ug-closure}b, which gives rise to a corresponding
shift in momentum.)

A remarkable feature which distinguishes the entanglement spectrum from
the one obtained for the RVB
state~\cite{poilblanc:rvb-boundaries,schuch:rvb-kagome} is that the
(possibly gapless) minimum of the dispersion in the half-integer spin
sector is at $k=\pm\tfrac\pi2$ rather than at $k=0$; this might serve as an
indicator to numerically distinguish the two phases.

\section{Conclusions}

In this paper, we have considered variational wavefunctions for kagome
Heisenberg antiferromagnets (HAFMs).  We have introduced semionic Resonating
Valence Bond (RVB) states and semionic dimer models, which are constructed
to be in the same phase as the double semion model.  We have discussed how
to parametrize the ground state manifold of this model, and have
subsequently used PEPS techniques based on the transfer operator to show
that the semionic RVB state is indeed in the same phase as the double
semion model.  We have also verified that the semionic RVB state does not
break spin $\mathrm{SU}(2)$ symmetry; it does however explicitly break the
symmetry of the lattice by construction. We have also computed the energy
w.r.t.\ kagome HAFM and found it to be not competitive with the
conventional RVB ansatz.

We have subsequently generalized semionic RVB states to semionic simplex
RVB states, which use next-nearest neighbor singlets to decrease the
energy of the ansatz.  We have found that with this optimization, the
energy of the semionic simplex RVB, $E_\mathrm{semRVB}^\mathrm{simplex} =
-0.4196$, outperforms the energy of the conventional simplex RVB with the
same optimization, while still being in the phase of the double semion
model, leaving open the possibility that the topological order of the
kagome HAFM is described by the double semion model.

We have also computed the entanglement properties of the semionic and
semionic simplex RVB.  We have found that both exhibit a $\mathbb Z_2$
correction to the topological entropy, and we have found that the
entanglement spectrum of the semionic RVB exhibits features which clearly
distinguish it from the entanglement spectrum of the conventional RVB
state; in particular, while both seem to be gapless, the minimum of the
dispersion is at $k=\tfrac\pi2$ for the semionic RVB, as opposed to $k=0$
for the conventional RVB.

Let us finally comment on the breaking of lattice (space group) symmetry
which naturally appears in the semionic RVB from the choice of a reference
configuration. It is not clear yet whether such a discrete symmetry
breaking is an essential feature of any RVB state with that kind of
topological order or whether it is also possible to construct
translation-invariant wavefunctions.  In any case, the coexistence of
so-called "Valence Bond Crystal" (VBC) order and topological order might
be relevant to the original kagome HAFM. Indeed, small bond modulations
seem to be ubiquitous in DMRG simulations on finite
clusters~\cite{yan:heisenberg-kagome,depenbrock:kagome-heisenberg-dmrg}
and whether those completely disappear in the thermodynamic limit is under
debate. In addition, Lanczos exact diagonalization of an effective quantum
dimer model provide evidence of competing VBC ground
states~\cite{poilblanc:kagome-qdm-vbc} including a $2\times 1$ "columnar"
VBC (CVBC) which possesses the same real-space structure as the semionic
RVB constructed in this work.  Note that a CVBC with coexisting
double-semionic topological order would give a degeneracy of $96$ ($24$ due to
space-group symmetry breaking times $4$ for topological order) of the
ground state space in the thermodynamic limit.  Interestingly, we expect
that this degeneracy might not be completely lifted in finite systems, as
the two semionic sectors should remain degenerate. Whether semionic VBC
phases can be realized in generalized quantum dimer
model~\cite{poilblanc:kagome-qdm-vbc} is left for further studies.

\emph{Note added:} During completion of this work, we learned
that Qi, Gu, and Yao~\cite{qi:semionic-dimer-models} had independently
introduced quantum dimer models with double-semion topological order in a
way similar to ours, and studied their corresponding Hamiltonians and the
nature of their excitations; and that Zaletel, Lu, and
Vishvanath~\cite{zaletel:private} have found general symmetry arguments
which suggest that it is impossible to construct a
translationally invariant double-semion state on the kagome lattice.

\subsection*{Acknowledgements}

M.I.\ and N.S.\ acknowledge funding by the Alexander von Humboldt
foundation, and computational resources provided by JARA-HPC via grants
jara0084 and jara0092. D.P.\ acknowledges fundings by the "Agence
Nationale de la Recherche" under grant No.\ ANR 2010 BLANC 0406-0.

\vspace*{0.5cm}

\onecolumngrid

\begin{center}
\rule{12cm}{0.03cm}
\end{center}

\vspace*{0.5cm}

\twocolumngrid

\appendix

\section{\label{sec:app:groundstates}
Ground state parametrization}

In this appendix, we briefly sketch how to show analytically that the
parametrization of the ground states of the double semion model in terms of
group elements and irreps of $\mathbb Z_4$ only gives rise to $4$ distinct
ground states, Eq.~(\ref{eq:topo-groundstates}).  All arguments work at
the level of a single column, and are given for the loop model (since the
RVB states constructed from it are always less distinguishable).

Let us first show that eight of the states vanish. To this end, consider a
column of the PEPS, and consider the case where the flux (the horizontal
string of $U_\phi$ in Fig.~\ref{fig:Ug-closure}b) is trivial,
$U_\phi=\openone$, $\phi=0$.  This implies that the color index is not flipped when
closing the boundary with $U_\phi$, and thus, the number of loops across
any vertical cut (and thus in particular at the left and right boundary)
is even.  On the other hand, the irreps $c=\pm i$ are supported
inside the subspace with an odd number of strings at the boundary [since
every string acquires a phase $i$ from $\eta$, Eq.~(\ref{eq:zeta-def})],
and thus, the states with $\phi=0$ and $c=\pm i$ vanish. The argument for
the other states goes analogously.

Second, let us show why pairs of boundary conditions describe the same
state. Again, consider the case with no flux, $U_\phi=\openone$, $\phi=0$.
and with trivial irrep $c=1$. Having $c=\pm 1$ is equivalent to a projection
$P_\pm\otimes \Pi_{ {0\mathrm{\,mod\,}4}} + P_\mp\otimes
\Pi_{2\mathrm{\,mod\,}4}$, where
$P_{+}$ $(P_{-}$) denotes the projection onto the subspace of 
states at one boundary which 
are invariant (change their sign) under flipping of all colors, and
$\Pi_{k\mathrm{\,mod\,}4}$ denotes the space of loop configurations with
$k\mathrm{\,mod\,}4$
loops at the boundary.Now consider $c=1$ and start with a column with no
loops at all: Flipping all colors at one side gives rise to a closed loop around
the column, which leads to no sign change.  For $c=-1$, the same procedure
flips the overall sign, which can however be compensated by a $U_\phi$
closure,
$\phi=2$, on a vertical bond, which corresponds to a $Z$ on the loop index
and thus exactly undoes the sign change, thus leaving the state invariant. 
Starting from this empty
loop configuration, we can now create any other loop configuration in the
$c=\pm1$ subspace by elementary moves, and it is straightforward to check
that all of these give the same result whether $c=1$ and $\phi=0$ or
$c=-1$ and $\phi=2$.  The argument for the other pairs goes again
analogously.

\section{\label{sec:app:numerical-optimization}
Numerical implementation}

\begin{figure}
\includegraphics[width=0.65\columnwidth]{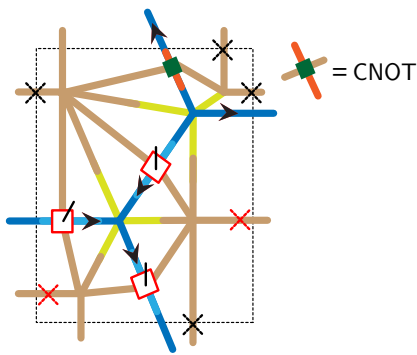} 
\caption{\label{fig:tn-optimizations}
Optimizations for the semionic RVB tensor network (see text).
}
\end{figure}

For the numerical implementation, we have used exact diagonalization of
the transfer operator using iterative eigensolvers and standard tensor
network methods; in particular, the transfer operator,
Fig.~\ref{fig:block-refconf-top}d, is applied to a vector by contracting
one tensor $\mathbb E$ after the other with the vector.  However, the
original tensor network still has a bond dimension $D=2\times 2\times 3 =
12$. Thus, we
first simplify the tensor network to reduce the bond dimension.  The
basic idea is to make use of the fact that we have a redundant
description, as we have both a color variable and a link variable (encoded
in the $E$ tensors). Thus, we can reduce the number of bonds used for
color variables, by re-computing the color variable every time a link is
crossed.  This is illustrated in Fig.~\ref{fig:tn-optimizations}:  We
start by removing all color indices which are marked with a black cross.
The color information is still passed by one bond in horizontal and one in
vertical direction, and is thus available in every tensor; the
construction of the tensor network (namely, the combined $D$ and $P$
tensor) ensures that the color index is flipped every time it crosses a
link. However, this is not true for the top right plaquette in the tensor;
this is resolved by adding a ``CNOT'' tensor which enforces the plaquette
colors to be equal or different depending on whether the link index is
$\{0,1\}$ or $2$, relative to the reference configuration.
A further improvement can be obtained by additionally removing the indices
marked by a red cross in all the tensors but one in a column; the
remaining tensor then passes the color index in the horizonal direction,
while it is passed vertically within each column. This leads to a further
reduction of the relevant horizontal bond dimension $D_h$ (the memory
requirement grows like $D_h^{2N_v}$).  In a final step, we consider the ket-bra
tensor $\mathbb E$ (Fig.~\ref{fig:block-refconf-top}e) obtained from the
optimized tensor in a $2\times 1$ unit cell and further optimize the bond
dimension using a singular value decomposition.

\section{\label{sec:app:dep-on-reference}
Different reference configurations}

\begin{figure}[t]
\includegraphics[width=0.95\columnwidth]{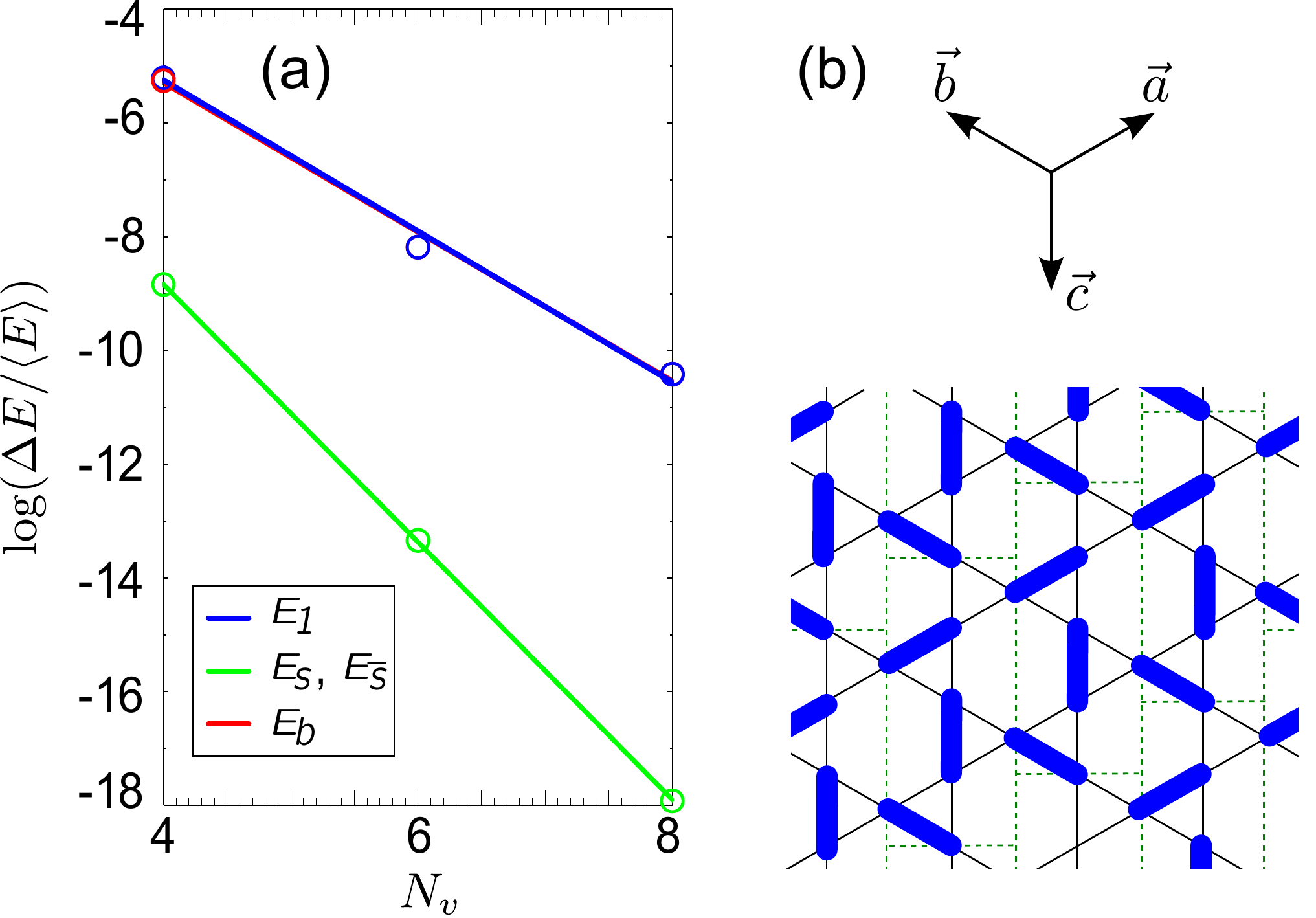}
\caption{
\label{fig:referenceconfig1}
\textbf{(a)} Energy difference between inequivalent reference
configurations.  The plot shows the difference of the average energy per
bond between semionic RVBs with two inequivalent reference configurations
as a function of $N_v$; we find that the energies converge exponentially
in $N_v$.
\textbf{(b)} The reference configuration of
Fig.~\ref{fig:block-refconf-top}b, with the blocking of
Fig.~\ref{fig:block-refconf-top}a (indicated by green dotted lines).  
The blocking gives rise to columnar order along
$\vec a$, which is left invariant by inversion, translation, and
reflection about $\vec a$, where the last transformation is incompatible
with the cylinder geometry.
}
\end{figure}

In this appendix, we briefly discuss the dependence of the energies on the
choice of reference configurations. In Sec.~\ref{sec:results}, we have
fixed a $6$-site unit cell, cf.~Fig.~\ref{fig:block-refconf-top}a.  Within
this unit cell, we can choose $8$ possible reference configurations (such
as the one in Fig.~\ref{fig:block-refconf-top}b).  We find
numerically that they form two groups of $4$ reference configurations each:
For any reference configuration within each group, the variational
energies are identical for any finite $N_v$
(though with a different symmetry breaking
pattern); on the other hand, reference configurations from different
groups give different energies, which however converge to the same value
in the thermodynamic limit $N_v\rightarrow\infty$ as shown in
Fig.~\ref{fig:referenceconfig1}a for the HAFM Hamiltonian. As we explain in
the following, this can be understood from the symmetries of the infinite
kagome lattice vs.\
the symmetries of the (YC) kagome lattice on a cylinder.

Given our choice of the $6$-site unit cell, Fig.~\ref{fig:block-refconf-top}a, each
reference configuration leads to a dimer covering of the kagome lattice
with columnar order, illustrated in Fig.~\ref{fig:referenceconfig1}b
for the reference configuration of
Fig.~\ref{fig:block-refconf-top}b. Here, the dimers arrange in columns
along the $\vec a$ axis; indeed, one can easily see that the orientation of the
columns originates from the choice of our $6$-site unit cell. There are
$3$ symmetry transformations which preserve the columnar dimer order along
$\vec a$: Inversion, translation, and reflection about the $\vec a$ axis, giving
rise to $8$ different dimer configurations with columnar order along $\vec a$.
While inversion and translation respect the symmetry of the finite
cylinder, and thus give rise to the same energy, reflection about the
$\vec a$ axis does not, which explains the different energies for finite
$N_v$.  On the other hand, it is of course a symmetry of the infinite lattice,
which is why the energies converge for $N_v\rightarrow\infty$.

\begin{figure}
\includegraphics[width=0.9\columnwidth]{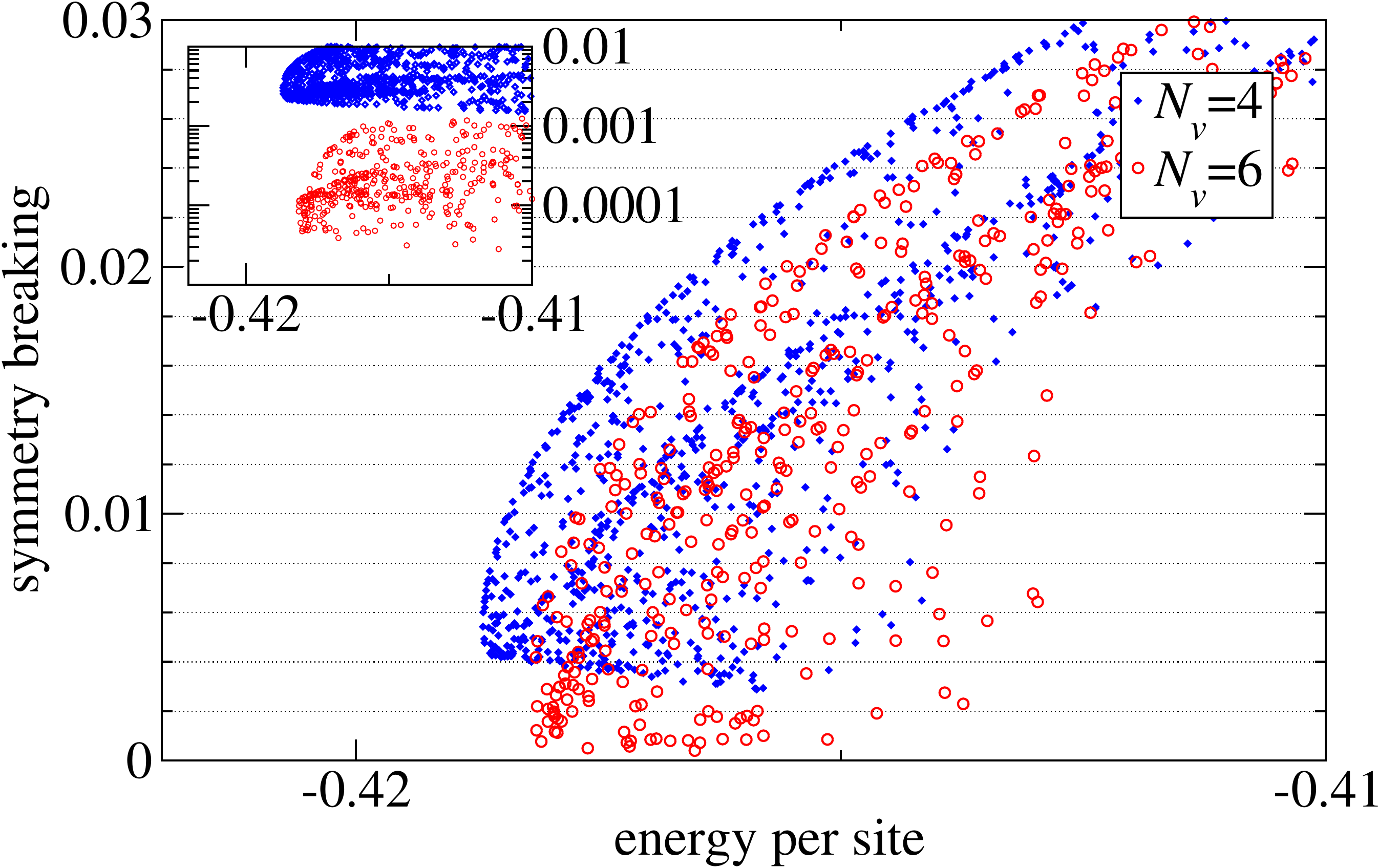}
\caption{\label{fig:symbreaking-normalrvb}
Symmetry breaking and finite size effects (inset) vs.\ variational energy
for the conventional simplex RVB, for randomly chosen $\alpha$ and $\beta$
around the optimal value.  The quantities plotted are the same as in
Fig.~\ref{fig:variations-vs-energy}.
}
\end{figure}

There are also two other possible orientations of the columnar order along
$\vec b$ or $\vec c$. Columnar order along $\vec b$ is related to $\vec
a$ by a reflection about the $\vec c$ axis, which respects
the symmetry of the cylinder, and thus gives identical energies as before
(though it corresponds to a shifted blocking of the PEPS as compared to 
Fig.~\ref{fig:square-blocking}a); thus, each of the energies belongs in
fact to a set of $8$ reference dimer coverings ($4$ along $\vec a$ and $4$
along $\vec b$). Columnar order along $\vec c$, on the other hand, is not
related by a cylinder symmetry to the other
cases and requires a horizontally oriented $6$-site unit cell; it
therefore gives rise to a third group of $8$ equivalent reference dimer
coverings which we expect to have a yet different energy for finite $N_v$,
and to converge again to the same value for $N_v\rightarrow\infty$.

\vspace*{0.5cm}

\section{\label{sec:app:symbreaking-normal-rvb}
Symmetry breaking in the conventional simplex RVB state}

In Fig.~\ref{fig:symbreaking-normalrvb} we provide the data on symmetry
breaking and finite size effects vs.\ variational energy for the
conventional simplex RVB,~\cite{poilblanc:simplex-rvb} in analogy to
Fig.~\ref{fig:variations-vs-energy}. We see that both the symmetry
breaking and finite size effects are much smaller than in the case of the
semionic simplex RVB, suggesting that symmetry breaking vanishes in
the thermodynamic limit.

\end{document}